\documentclass[pra,twocolumn,showpacs,superscriptaddress]{revtex4-1}

\usepackage{multirow,eurosym,amssymb,amsfonts,graphicx,color,bm,float,verbatim}
\expandafter\let\csname equation*\endcsname\relax
\expandafter\let\csname endequation*\endcsname\relax
\usepackage{amsmath}
\def\openone{\leavevmode\hbox{\small1\kern-3.8pt\normalsize1}}

\newcommand{\ket}[1]{|#1\rangle}

\begin{document}

\title{Generating and verifying graph states for fault-tolerant topological
measurement-based quantum computing in 2D optical lattices}

\author{Jaewoo Joo}
\address{Quantum Information Science, School of Physics and Astronomy, University of Leeds, Leeds LS2 9JT, U.K.}

\author{Emilio Alba}
\address{ Instituto de F\'{i}sica Fundamental, Consejo
Superior de Investigaciones Cient\'{i}ficas, Calle Serrano 113b,
Madrid E-28006, Spain}

\author{Juan Jos\'{e} Garc\'{i}a-Ripoll}
\address{ Instituto de F\'{i}sica Fundamental, Consejo
Superior de Investigaciones Cient\'{i}ficas, Calle Serrano 113b,
Madrid E-28006, Spain}

\author{Timothy P. Spiller}
\address{Quantum Information Science, School of Physics and
Astronomy, University of Leeds, Leeds LS2 9JT, U.K.}

\begin{abstract}
We propose two schemes for implementing graph states useful for fault-tolerant topological measurement-based quantum computation in 2D optical lattices. We show that bilayer cluster and surface code states can be created by global single-row and controlled-Z operations. The schemes benefit from the accessibility of atom addressing on 2D optical lattices and the existence of an efficient verification protocol which allows us to ensure the experimental feasibility of measuring the fidelity of the system against the ideal graph state. The simulation results show potential for a physical realization toward fault-tolerant measurement-based quantum computation against dephasing and unitary phase errors in optical lattices.
\end{abstract}

\maketitle

\section{Introduction} 
Measurement-based quantum computation (MBQC) has provided a new paradigm of quantum computation for a decade \cite{Oneway}.
This model requires a highly entangled state initially prepared by two-qubit entangling gates (i.e., controlled-Z (CZ) gate) before the beginning of information processing. With the resource state, a sequence of single-qubit measurements is solely required to simulate any quantum algorithm for MBQC. As an example of the resource states, a two-dimensional (2D) cluster state \cite{Cluster} is the well-known (and the first-known) graph state associated with mathematical graphs \cite{graphstate01}. Note that vertices and edges represent qubits and CZ gates, respectively, in graph representation. To build a 2D cluster state, two steps are in general required. All qubits in the lattice are first initialized in state $|+\rangle = (|0\rangle + |1\rangle)/\sqrt{2}$ in a 2D lattice system and CZ operations are then globally performed between every qubit and its four neighbouring qubits. Even though many realization schemes have been proposed in various physical systems, a simple approach to MBQC would be susceptible to errors and decoherence \cite{James09,MBQC_propose}. For example, even if a 2D cluster state is ideally prepared, a small error during the single-qubit measurement procedure could be propagated to other qubits in the rest of the entangled system, with the accumulated errors resulting in incorrect outcomes of the quantum computation. The long term goal would therefore be to perform fault-tolerant MBQC against the effects of errors/decoherence for practical application. 

Several fault-tolerant schemes are in principle achievable for MBQC. The construction of concatenated logical cluster states using quantum error-correcting codes \cite{Joo-Fujii} and the engineering of thermal ground states of high-spin systems \cite{Wei} have been proposed as theoretical platforms for fault-tolerant MBQC. However, the implementation of these approaches would be difficult to create such complicated entanglement structures in natural physical systems. The alternative of performing fault-tolerance in MBQC is to create topologically entangled states such as 3D (as opposed to 2D) cluster states and MBQC can be fault-tolerantly performed on the 3D clusters via a sequential single-qubit measurements \cite{Raussendorf06,MunroNJP09,MoreMunro}. It is also known that the 3D cluster states can be transformed to surface codes (also called toric codes of Kitaev \cite{Kitaev03}), which provide the known highest error threshold for universal fault-tolerant quantum computation \cite{Raussendorf07,Fowler}. Surface code states are a class of stabilizer quantum codes associated with lattices and provide an excellent platform for fault-tolerant quantum information tasks using the property of topological systems \cite{Kitaev03}. Qubits are located on the edges of the 2D lattice and stabilizer operators are defined in terms of four Pauli operators. The first in-principle demonstration has been very recently performed in photonic graph states \cite{Pan11} and its practical approaches have been also investigated for quantum memory and quantum communication in surface code lattices \cite{Jiannis}.

Optical lattices are a powerful platform in which to build and operate a MBQC protocol \cite{Bloch08}. On the one hand, they are naturally suited to create initialized and entangled states with a large number of neutral cold atoms efficiently. On the other hand, recent experiments have demonstrated the ability of manipulating and measuring individual atoms \cite{Kuhr11} and of performing entangling operations between neighbouring atoms in 2D optical lattices \cite{Soderberg09}. However, the significant issues still exist for the implementation of resource states toward fault-tolerant MBQC in optical lattices. One of the major issues is that an atomic 3D cluster state can be in general built in a 3D {\em spatial} configuration of optical lattices \cite{Whaley08}. This has a significant hurdle because of site accessibility and lack of single-atom addressability. Another issue is that it is very difficult to check and verify the entanglement in the prepared entangled/graph states \cite{Plenio11} when relying on tomography by individual qubit measurements. Thus, an efficient verification scheme is also an important and urgent matter in MBQC based on optical lattices. 

We here present two theoretical implementation schemes that provide potential for fault-tolerant topological MBQC in 2D optical lattices. 
A scheme of embedding a 3D (in state space) cluster state in 2D physical structure is first investigated (inspired by Ref.~\cite{James09}) and the other shows that surface code states can be directly created with the help of global single-atom measurements in one of two optical attices. Note that we use a clear distinction between the spatial dimensionality in real physical space and the effective spatial dimension of the quantum resource constructed. The two schemes share the same experimental setup: two independent square lattices from a diffraction image. In line with the experimental setup proposed in Ref.~\cite{Soderberg09}, the species of two atoms can be considered such as ${}^6$Li and ${}^{133}$Cs (denoted by red dotted and blue solid circles in Fig.~\ref{fig:GlobalH01}). In addition to the advantage of the 2D configuration, we provide the addressability of the individual atoms and an efficient verification scheme based on a few global single-qubit measurements.

This paper is organised as following.
In Section \ref{sec:setup}, we provide the background of the specific realization with state-of-the-art optical lattice techniques. Particularly, we present a method of single-row Hadamard operations at every second row of 2D optical lattices and the tool is used in both implementation schemes of creating 3D cluster and surface code states in Section \ref{sec:3}. The simulation protocol and its results are given in the presence of errors in Section \ref{sec:4} and a summary and remarks are finally presented in Section \ref{sec:5}. 
 
\section{Background: optical lattice setup}\label{sec:setup}  
\begin{figure}[t]
\hspace{-1.5cm}
\includegraphics[height=6cm,angle=-90]{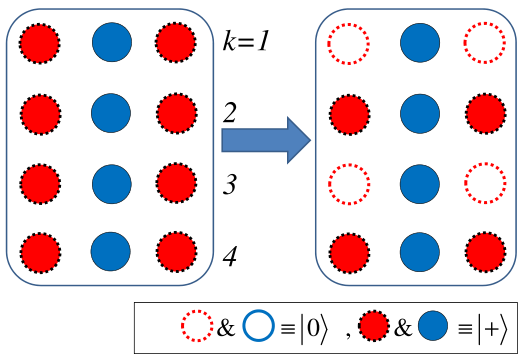}
\caption{(Color online) The schematics of global Hadamard operation at every second row in red (dotted) atoms (e.g.,$^{6}$Li).}
\label{fig:GlobalH01}
\end{figure}

The natural lattice structure of graph states suggests that optical lattices can be used as an excellent experimental platform \cite{Bloch08}. Here we will consider a superposition of two different bidimensional optical lattices, each of which hosts a different (alkaline) atomic species. The two species can be simultaneously trapped in different coplanar lattices by focusing a diffraction image of the lattice onto the focal plane of a large aperture lens~ \cite{Soderberg09}. Qubits are encoded in hyperfine ground state levels of the atoms. Optical lattices are then very suitable for building graph states due to the ability to perform global two-qubit entangling interactions. In particular, controlled collisions \cite{Jaksch98} and engineered interaction \cite{Soderberg09,Yelin12,Alba10} provide excellent examples of desired two-qubit interactions that can be manipulated globally across a lattice. We emphasize the experimental setup in Ref.~\cite{Soderberg09} both for implementing CZ gates by moving atoms to overlap the different species and then applying rf pulses and for building a bipartite graph state using two different species of the atoms in 2D optical lattices. Then, we obtain a CZ operation in different sublattices such that $U_{\rm CZ}^{j,k}=\exp (- i {\pi \over 4} \sigma^z_j \sigma^z_k)$ ($\sigma$ is a Pauli operator). 

Due to the requirement of single-qubit measurements in MBQC, significant experimental progress has been made toward single-atom addressing (e.g., using pointer atoms, laser interference patterns, microwave transitions, etc.)~\cite{SAA-theory,SAA-theory2,Kuhr11,magnetic-SQA,magnetic4} for creating resource states and performing the measurement protocol. measuring the atomic states could be optimally implemented in the prepared resource states using single-atom fluorescence imaging. For example, measuring the atomic states could be optimally implemented in the prepared resource states using single-atom fluorescence imaging.

One of the key advantages of the proposed setup is that we do not specifically need single-site resolution in order to prepare/verify the resource states because of the bipartite architecture, while providing enough physical space for single-qubit measurement with the current optical technology. Moreover, no edge (no CZ gate) is needed between the same atomic species. This architecture has the additional advantage  that a lower bound on the fidelity to the desired graph state can be computed by globally measuring one spin observable in each atomic species \cite{Alba10}. This greatly simplifies the task of verifying whether the experimental procedure has been successful to create useful graph states for MBQC.

To construct desired graph states, a global single-row Hadamard operation is required at every second row. As shown in Fig.~\ref{fig:GlobalH01}, the global (pseudo) Hadamard operation is inspired by Ref.~\cite{Viv02} and given by 
\begin{eqnarray}
H_{pseudo} (k)=R_z({\pi\over 4})R_y(k\pi)R_z({\pi\over 4}) = (-i)^{k} H^{k-1},
\label{Hadamard}
\end{eqnarray}
for $H$ is the Hadamard operator and $k$ indicates the $k$-th row in a 2D optical lattice ($R_M (\theta) = \exp [-i\theta \sigma^{M}/2]$ for $M=x,y,z$). This global Hadamard gate can be implemented by applying two additional Raman lasers on a standing wave configuration whose period doubles that of the individual sublattices. As shown in Fig.~\ref{fig:GlobalH01}, acting solely on the ``red'' Li sites, the Raman beams configuration would implement a Hadamard gate on every other row, creating an alternation between the states $\ket{0}$ and $\ket{+}$. 

\section{Two schemes}
\label{sec:3}
\begin{figure}[t]
\hspace{-1.5cm}
\includegraphics[height=7cm,angle=-90]{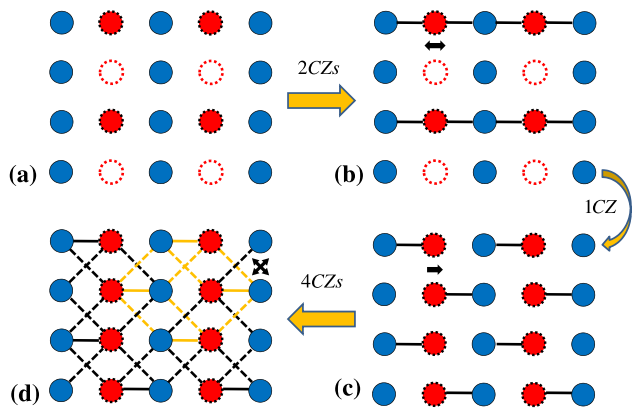}
\caption{(Color online) To build 3D cluster states, 7 CZ and 2
Hadamard operations are globally required in 2D optical lattices. Solid lines indicate in the CZ gates in horizontal/vertical direction while dashed lines do CZ operations at diagonal directions (small black arrows show the direction of moving atoms). In (d), the colored yellow edges
present a unit cell of 3D cluster states in the lattices.}
\label{fig:01}
\end{figure}

We now present two theoretical implementation schemes that provide potential for fault-tolerant MBQC built in 2D optical lattices. In scheme (i) we demonstrate how to construct 3D (in state space) cluster states in the 2D lattices. In scheme (ii) we show how to create a surface code state from a 2D graph state directly. Both schemes are designed in the same experimental setup and operations: the global pseudo Hadamard and patterned CZ operations. We assume that both schemes start with patterned non-entangled states (see Figs. \ref{fig:01}(a) and \ref{fig:02}(a)), which can be prepared by applying the global Hadamard operations in advance.

\subsection{Scheme (i): 3D cluster states }
The first implementation scheme, for building a bilayer (effective) 3D cluster state, is inspired by Ref.~\cite{James09}. In contrast to the ion trap scheme, we provide a verification protocol that is a simple but powerful tool to check the multipartite entanglement in a 2D atomic lattice. The construction of the 3D cluster state in (spatial) 2D optical lattices requires seven global CZ operations overall, along with the patterned Hadamard operations already described. The detailed operational sequence is depicted in Fig.~\ref{fig:01} and note that small arrows indicate the direction of moving optical lattices to perform CZ operations. a) The starting point is the preparation of optical lattices in the patterned atomic states, and red atoms of even number rows are initialized in state $\ket{0}$ by global Hadamard operations. b) Once the 1D cluster states are made by 2 global CZ operations at the odd number rows while no entanglement exists between atoms in the even number rows. An additional patterned Hadamard operation is then performed at every second row of red (Li) atoms. c) An additional global CZ operation in a direction (e.g., CZ from red to blue) is operated to obtain the pair-wise patterned entangled state (see Fig.~\ref{fig:01}(c)). Note that some CZ edges are vanished at the odd number columns due to $(U_{\rm CZ})^2 = \openone$. (d) We perform four CZ operations in the diagonal directions. The final outcome is equal to a bilayer 3D cluster states in a 2D optical lattice system. Due to the physical 2D structure, our scheme provides a measurement capability for the full atomic 3D cluster state (e.g., with lasers directed from the other spatial dimension orthogonal to the 2D lattice). In principle, the layers in the 3D cluster could be extended to multi-layers by adding different species of atoms in the 2D optical lattice.

\subsection{ Scheme (ii): surface code states } 
\begin{figure}[t]
\centering
\hspace{-1.5cm}
\includegraphics[height=7cm,angle=-90]{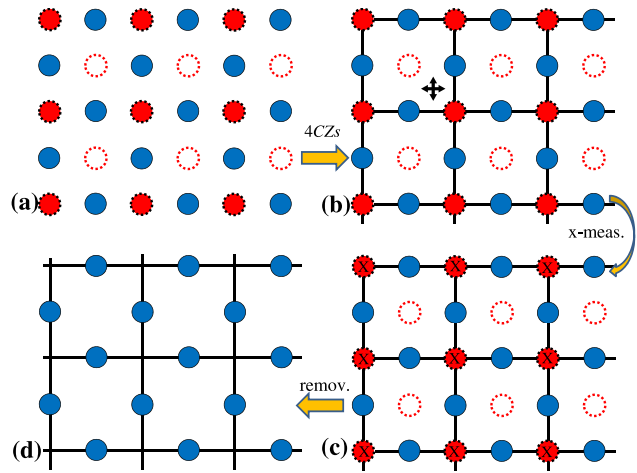}
\vspace{1cm} \caption{(Color online) How to build a surface code directly in 2D optical lattices. The surface code is directly created by Pauli $\sigma^{x}$ measurements (denoted by X) at all red (Li) atoms in (c) \cite{Raussendorf07,Soderberg09}. Then, red atoms are removed by turning off their optical lattices.} \label{fig:02}
\end{figure}
In the second implementation scheme, we demonstrate that the surface code can be directly created by local $\sigma^{x}$ measurements from a graph (2D cluster-like) state. It is even more economical (in terms of global gates) to build the desired entangled state that embeds the surface code. a) The patterned state is the initialized state as shown in Fig.~\ref{fig:02}(a). b) After four CZ operations are performed between dotted red (Li) and solid blue (Cs) atoms (see small black arrows), the intermediate state has patterned entanglement as a bipartite graph state, which can be used for performing a verification protocol (see details in Section \ref{sec:4}). c) Taking the advantage of the interlocking different species of optical lattices, we globally perform $\sigma^{x}$ measurements only on all red (Li) atoms. The measurement outcomes can provide us the information of qubits in optical lattices \cite{Jiannis}. d) After the global Pauli $\sigma^{x}$ measurements, all the dotted atoms are dropped by turning off their optical lattice. The resultant state is therefore equivalent to the surface code encoded state as shown in Fig.~\ref{fig:02}(d).

\section{Simulations}
\label{sec:4}
Building the graph states proposed in the two schemes we have presented requires several state-of-the-art techniques, such as hyperfine state initialization and global entangling 2-qubit gates by overlapping pairs of atoms. This section presents a practical protocol for calculating a lower bound to the fidelity of our state to the desired bipartite graph state, and show numerical results to demonstrate how errors might affect the fidelity of the system. This verification protocol not only acts as the measure of perfectness for the desired MBQC resource state but also detects localized errors in the setup. Thus, it provides a very powerful and useful tool for calibrating the perfectness of creating prepared states in experiments.

\subsection{Verification protocol }
Verifying the success of the experimental protocol which creates
the bipartite graph state (e.g., Fig.~\ref{fig:01}(d) and~\ref{fig:02}(b)) is a
powerful tool to ensure MBQC can be performed in the designed
setup. It has been shown that the connectivity of graph states in
2D optical lattices can be verified by solely using the
performance of $\sigma^{x}$ and $\sigma^{z}$ measurements
\cite{Alba10}. This is so because the graph state is the eigenstate 
of the set of stabilizer operators $S_i= \sigma^x_i \prod_{j \in \mathcal{N}(i)}\sigma^z_j$ with eigenvalue $+1$, where
$\mathcal{N}(i)$ stands for all nearest neighbours of the lattice
site $i$. Therefore $P=\prod_i (1+S_i)/2$ is a projector which
directly measures the fidelity of any give state to a graph state.

A remarkable feature of the proposed schemes is that the
resulting graph states are bipartite. For example, they can be represented by a graph composed by two disjoint unconnected sets of vertices~\cite{Hein06}. This property is enough to show that the fidelity of the experimental state to a graph state can be computed with only two sets of operators~\cite{Guhne09}. We shall denote by $A$ and $B$ two sets of lattice sites (e.g., $A$=Li and $B$=Cs). These sets are to be complementary ($A \cup B$ is the whole lattice) and disjoint ($A \cap B = \emptyset$). It is possible to simultaneously measure $\sigma^x_i$ for all $i \in A$ and $\sigma^z_j$ for all $j \in B$ since all these Pauli operators commute with each other. This also holds for the simultaneous measurement of $\sigma^z_i$ for all $i \in A$ and $\sigma^x_j$ for all $j \in B$. These single-site measurements suffice to compute
\begin{eqnarray}
\label{projectors}
\langle P_M \rangle &=& \left\langle \prod_{i \in M} \frac{1}{2} \left( \openone + \sigma^x_i \prod_{j \in \mathcal{N}(i)}\sigma^z_j \right) \right\rangle
\end{eqnarray}
where $\mathcal{N}(i)$ stands for all nearest neighbours of $i$ in
the graph associated to the graph state ($M=A,B$); this property is verified in every bipartite lattice by construction of $P_M$. Each of the
two projectors can be evaluated with a single experimental setup,
and the fidelity of the system to the ideal graph state can be
bounded by 
\begin{eqnarray}
\mathcal{F} \geq \langle P_A \rangle + \langle P_B \rangle -1,
\label{Fidelity01}
\end{eqnarray}
\cite{Alba10} and thus needs only two experimental setups~\cite{Guhne09}.
Moreover, this protocol allows for detection of high-fidelity
regions in the presence of defects~\cite{Alba10}.

\subsection{Simulation results}
\begin{figure}[t]
\centering \hspace{-1.5cm}
\includegraphics[height=7cm,angle=-90]{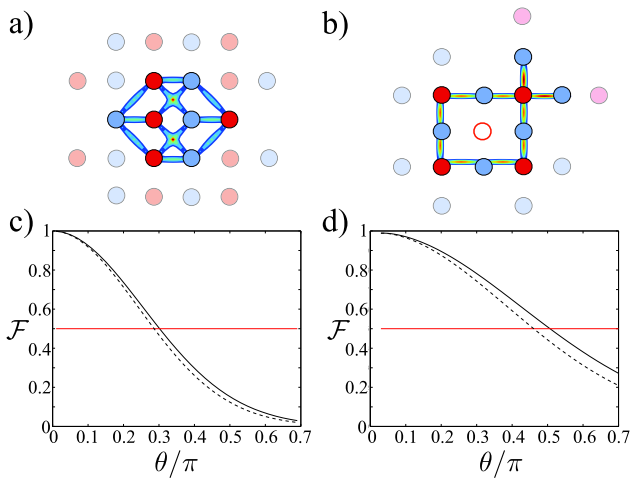}
\vspace{1.0cm}
\caption{(Color online) a-b) Simulation of the application of the
2-qubit verification protocol in the basic blocks shown in Fig.~\ref{fig:01} (d) and Fig.~\ref{fig:02} (b) respectively, where the simulations are computed with a fixed dephasing error parameter $\theta=\pi/5$. The color links show a positive result of the 2-qubit entanglement witness. Shaded atoms represent the border of the unit blocks, which has to be measured in order to compute the local fidelity. (c-d) Values of the fidelity to the graph state as a function of $\theta$ for the 3D cubic block (solid) and the surface-code block (dashed) for dephasing (c) and unitary CZ (Ising) (d) errors. The horizontal line shows the minimum value for the observable to show genuine multipartite entanglement. Notice that the plots are roughly similar for both blocks due to the similar size of the blocks considered.} \label{Sim}
\end{figure}

We now show simulations of applying the verification protocol to
the basic building blocks of our bipartite graph states in
Fig.~\ref{fig:01}(d) and Fig.~\ref{fig:02}(b). In order to apply the verification protocol to our schemes, we only need to
specify the disjoint sets $A$ and $B$ for each case. In the 3D cluster state of Fig.~\ref{fig:01}(d), these sets have to be
chosen as the even and odd columns of the lattice. It is remarkable that such a simple
pattern works for a projection of an effective cubic lattice. The
cluster state of Fig.~\ref{fig:02}(b), which is an intermediate
step towards the surface code, can be verified even more easily,
since the disjoint sets are the set of Li and Cs atoms. In in
Fig.~\ref{Sim}(a)-(b), shaded atoms represent the border of the
unit blocks, which has to be measured to compute the local
fidelity.

Several sources of experimental error can be quantified in
simulating the application of the protocol to a unit
block. We mainly focus on numerical simulations considering 
two main sources of errors from case 1: inaccurate qubit initialization~\cite{Alba10} and case 2: unitary phase errors in global CZ (Ising) operations ~\cite{David08} such as 
\begin{eqnarray}
\rho_{j} &\rightarrow& \int d \theta \exp \left (- i \sigma_z \theta_{j} \right) \rho_{j} \exp \left ( i \sigma_z \theta'_{j} \right), 
\label{dephasing01} \\
U_{\rm CZ}^{j,k} &\rightarrow& U_{\rm CZ}^{j,k}exp(i \theta_{j,k}\sigma_z^j\sigma_z^k),
\label{unitary01} 
\end{eqnarray}
where $j,k$ are the site numbers and an error angle is $\theta$.

In case 1, atom losses (or vacancies in optical lattices) could result in undesirable graph states and a simple source of errors could come from inaccurate qubit initialization and qubit is therefore subject to a dephasing map which can be quantified in terms of a dephasing angle $\theta$ in Eq.~(\ref{dephasing01})~\cite{Alba10}. In case 2, the actual phases in CZ operations are likely to be of the form $\pi + \theta$, with a small unknown Ising error $\theta$ \cite{David08}.  We therefore parametrize both decoherence channels by a parameter $\theta$. Theis parameter quantifies the deviation from the ideal pulse times or beam energies.  A typical error distribution, which we will assume in our simulations, is to take $\theta$ as a uniformly distributed random variable within the interval $[-\theta',\theta']$.

In Fig.~\ref{Sim}(a) and (b), we plot the two-qubit entanglement witness after applying a local dephasing map with a fixed (and moderately low) value of $\theta' = \pi/5$. Note that red color indicates the high fidelity and the links in the figures show that each pair of qubits is close to a highly entangled  state. Fig.~\ref{Sim}(c) and (d), shows the simulated bipartite bound of the local fidelity in Eq.~(\ref{Fidelity01}) for both proposed schemes after applying each of the considered decoherence maps. Particularly, an error parameter $\theta$ lower than $\pi/20$ is enough to obtain a fidelity above $0.98$ in both models,showing potential to reach the error-threshold for fault-tolerant MBQC.

\section{Summary and remarks}
\label{sec:5}
In summary, we have studied the implementation of useful
graph states for fault-tolerant MBQC in optical lattices. Only a global
single-row and CZ operations are required to build the two
graph states (3D bilayer cluster and surface code states). The creation of a 3D cluster state requires
7 CZ and 2 patterned Hadamard operations while that of a surface code state does 4 CZ operations with the help of local $\sigma^x$ measurements on one species of the atoms. According to the simulation results, The fidelity between ideal and prepared graph states
could overcome error-thresholds in both implementation schemes \cite{Raussendorf07,Fowler}.

The implementation approach has three main features. First, we propose to take
advantage of and utilise a practical approach for the creation of bipartite graph states,
that has recently been demonstrated in experiments with
 optical lattices \cite{Soderberg09}. Second, the
implementation scheme of 3D clusters offers us atomic
addressability because bilayer cluster states can be constructed within the geometry of 2D optical lattices. Finally, the bipartite graph states in 2D optical lattices provide a unique and efficient verification protocol, to
guarantee the fidelity of the resulting states before the
performance of fault-tolerant MBQC. In addition, the proposed schemes could therefore be useful for a practical approach in not only quantum computation but also quantum communication and quantum memory \cite{Jiannis}.

\section{Acknowledgements}
We acknowledge J. K. Pachos for useful discussions and financial support from the European Commission of the European Union under the FP7 Integrated Project Q-ESSENCE, the Spanish MICINN Project FIS2009-10061, FPU No.AP 2009-1761 and CAM research consortium QUITEMAD S2009-ESP-1594.

{}

\end{document}